\documentclass[structabstract]{aa}
\usepackage{natbib}			% BibTeX style bibliography
\usepackage{amsmath}			% Mathematic formulaes
\usepackage{graphicx} 			% Pictures and images
\usepackage[varg]{txfonts}		% A&A is printed using the Postscript TX Times-fonts.
\usepackage[english]{babel}		% Language is english
\usepackage[normalem]{ulem} 		% for striketrough type effects
\usepackage{paralist}			% lists
\usepackage{multirow}
\usepackage{hyperref}             
\hypersetup{backref=true, pagebackref=true, hyperindex=true, breaklinks=true,colorlinks=true,urlcolor=blue, linkcolor=blue,  citecolor=blue,pagecolor=red, bookmarks=true, bookmarksopen=true}
\usepackage{booktabs}
\begin{document}
  \title{On the stellar clustering and architecture of planetary systems}

  \author{V.~Adibekyan\inst{1,2} \and
          N.~C.~Santos \inst{1,2} \and
          O.~D.~S.~Demangeon \inst{1,2} \and
          J.~P.~Faria \inst{1,2} \and
   	  S.~C.~C.~Barros \inst{1,2} \and
    	  M.~Oshagh \inst{4,5} \and
   	  P.~Figueira\inst{3,1} \and
   	  E.~Delgado~Mena \inst{1} \and
          S.~G.~Sousa \inst{1} \and 
     	  G.~Israelian \inst{4,5} \and
    	  T.~Campante \inst{1,2} \and
    	  A.~A.~Hakobyan\inst{6}
 }

  \institute{
  	  Instituto de Astrof\'isica e Ci\^encias do Espa\c{c}o, Universidade do Porto, CAUP, Rua das Estrelas, 4150-762 Porto, Portugal \\
  	  \email{vadibekyan@astro.up.pt} \and
  	  Departamento de F\'{\i}sica e Astronomia, Faculdade de Ci\^encias, Universidade do Porto, Rua do Campo  Alegre, 4169-007 Porto, Portugal \and
   	  European Southern Observatory, Alonso de Córdova 3107, Vitacura, Región Metropolitana, Chile \and
  	  Instituto de Astrof\'{i}sica de Canarias, E-38205 La Laguna, Tenerife, Spain \and           Departamento de Astrof\`{i}sica, Universidad de La Laguna, E-38206 La Laguna, Tenerife, Spain \and
   	  Center for Cosmology and Astrophysics, Alikhanian National Science Laboratory, 2 Alikhanian Brothers Str., 0036 Yerevan, Armenia
  	  }

  \date{Received date / Accepted date }
%----------------------------------------------------------------------------------------
%	Abstract
%----------------------------------------------------------------------------------------
  \abstract
  {Revealing  the mechanisms shaping the architecture of planetary systems is crucial for our understanding of their  formation and evolution. In this context, it has been recently proposed that stellar clustering might be the key in shaping the orbital architecture of exoplanets.}
  {The main goal of this work is to explore the factors that shape the orbits of planets.}
  {We perform different statistical tests to compare the properties of planets and their host stars associated with different stellar environments.}
  {We used a homogeneous sample of relatively young FGK dwarf stars with RV detected planets and tested the hypothesis that their association to  phase space (position-velocity) over-densities (`cluster' stars) and under-densities (`field' stars) impacts the orbital periods of planets. When controlling for the host star properties, on a sample of 52 planets orbiting around ‘cluster’ stars and 15 planets orbiting around 'field' star,  we found no significant difference in the period distribution of planets orbiting these two populations of stars. By considering an extended sample of 73 planets orbiting around 'cluster' stars and 25 planets orbiting 'field' stars, a significant different in the planetary period distributions emerged. However, the hosts associated to stellar under-densities appeared to be significantly older than their 'cluster' counterparts. This did not allow us to conclude whether the planetary architecture is related to age, environment, or both.  We further studied a sample of planets orbiting `cluster' stars to study the mechanism responsible for the shaping of orbits of planets in similar environments. We could not identify a parameter that can unambiguously be responsible for the orbital architecture of  massive planets, perhaps, indicating the complexity of the issue.}
  {Increased number of planets in clusters and in over-density environments will help to build large and unbiased samples which will then allow to better understand the dominant processes shaping the orbits of planets.}
  

%----------------------------------------------------------------------------------
%	Title
%----------------------------------------------------------------------------------

  \maketitle
%---------------------------------
   
%  -------------------------------------------------
%	Introduction
%----------------------------------------------------------------------------------
\section{Introduction}					\label{sec:intro}

Understanding the mechanisms shaping the architecture of planetary systems is crucial to complete the picture of planet formation and evolution \citep[e.g.][]{Winn-15, Hatzes-16}. Among the many open questions in this field, it is of particular interest to understand the origin of hot Jupiters \citep[HJs,][]{Dawson-18} - short period giant planets\footnote{The definition of HJs in terms of upper limit of orbital period (or semi-major axis) and lower limit in planetary mass (or radius) varies in the literature.}. Several mechanisms are proposed to explain the presence of these massive planets at very close distances to their host stars: in-situ formation, disk migration, high-eccentricity tidal migration, and dynamical perturbations by stellar fly-bys in open clusters. Although, a combination of these mechanisms might be needed to explain the observational properties of HJs and their hosts stars \citep{Dawson-18}, it was very recently suggested that the short periods of HJs originate from environmental perturbations \citep[][hereafter, W20 ]{Winter-20}.

To study the possible link between stellar clustering and architecture of planetary systems, \citet{Winter-20} estimated the probability ($P_{\mathrm{high}}$) that a planet host star belongs to over-  or under-densities  in the position-velocity phase space. The authors determined and made publicly available the $P_{\mathrm{high}}$ values for more than 1500 exoplanet host stars for which radial velocities were available in Gaia DR2. Stars with $P_{\mathrm{high}} > 0.84$ were considered as potential members of co-moving groups (over-density or `cluster'  stars) and stars with $P_{\mathrm{high}} < 0.16$ as `field' stars. Based on this database, they reached to two important conclusions: Planets orbiting stars associated with over-densities have significantly shorter orbital periods than those orbiting around `field' stars and that HJs predominantly exist around `cluster' stars. 

Given the importance of these findings and conclusions, in this manuscript we performed an independent analysis of their data but using homogeneously determined stellar parameters of the planet host stars from the SWEET-Cat \citep{Santos-13}. 

The manuscript is organized as follows: In Sect.~\ref{sect:sweet-cat} we first built a homogeneous sample trying to control different biases and then studied the period distribution of planets orbiting `cluster' and `field' stars. In Sect.~\ref{sect:W20_reversed_analysis} we studied the impact of different physical parameters on the orbital periods of planets associated with high-density stellar environments. We summarize our work in Sect.~\ref{sect:summary}.

\section{SWEET-Cat FGK dwarf RV sample} \label{sect:sweet-cat}

In order to confirm or refute the main findings of \citet{Winter-20} it is crucial to perform the analysis on an unbiased sample. For a discussion about the impact of different (potential) biases we refer the reader to W20.  In this section we  build a sample (based on the original full sample of W20) of RV detected\footnote{Transiting planets have only short periods and are not suitable for our analysis.} high-mass planets orbiting around  FGK dwarf stars for which homogeneously derived (see Sect.~\ref{sect:parameters}) stellar parameters exist in SWEET-Cat. We then perform a statistical analysis on this data using the AD test to study the impact of stellar clustering on the orbital periods of exoplanets. It is important to note that by restricting the sample to RV detected planets we did not remove the observational biases that this planet detection method suffers from. However, we try to minimize the impacts of different biases by applying further restrictions on the properties of planets and their hosts. Ideally, one would have to carefully model and correct for the detection biases to construct the actual period distributions of the planets. However, this would be extremely difficult since the planets of our sample come from different planet search programs carried out with different instruments, different observational strategies, and different detection biases.

In this analysis we focus only on massive planets with masses between 50~$M_{\oplus}$ and 4~$M_{\mathrm{jup}}$. The selected lower limit is the same as the one adopted in W20 for HJs. This semi-arbitrary limit \citep[see the discussion in][]{ Adibekyan-19} is considered to decrease the impact of planet detection limits (low-mass planets are difficult to detect in wide orbits) and also  the planet core-accretion  models predicted a minimum in the planetary mass-distribution at about 50~$M_{\oplus}$ mass \citep[e.g.][]{Mordasini-09}. Our choice of upper mass limit is motivated by the recent findings that the properties of stars hosting super-massive Jupiters ($M_{\mathrm{pl}} > 4~M_{\mathrm{jup}}$)  are different from those hosting lower-mass Jupiters which might suggest a different formation mechanisms \citep[e.g.][]{Santos-17, Adibekyan-19, Maldonado-19, Goda-19}. The range of effective temperature of the selected FGK stars is 4500 $< T_{\mathrm{eff}} <$ 6500 K. This is the range of temperatures for which SWEET-Cat provides most precise stellar parameters \cite[e.g.][]{Sousa-08}. From our sample we excluded the evolved stars ($\log g < 4.0$ dex) because their properties (for example their mass and metallicity) and the properties of their planets (for example the orbital periods) show different distributions when compared with the properties of planets orbiting around dwarfs \citep[e.g.][]{Adibekyan-13,Maldonado-13,Mortier-13}.

\begin{figure*}
\begin{center}
\begin{tabular}{cc}
 \includegraphics[angle=0,width=0.45\linewidth]{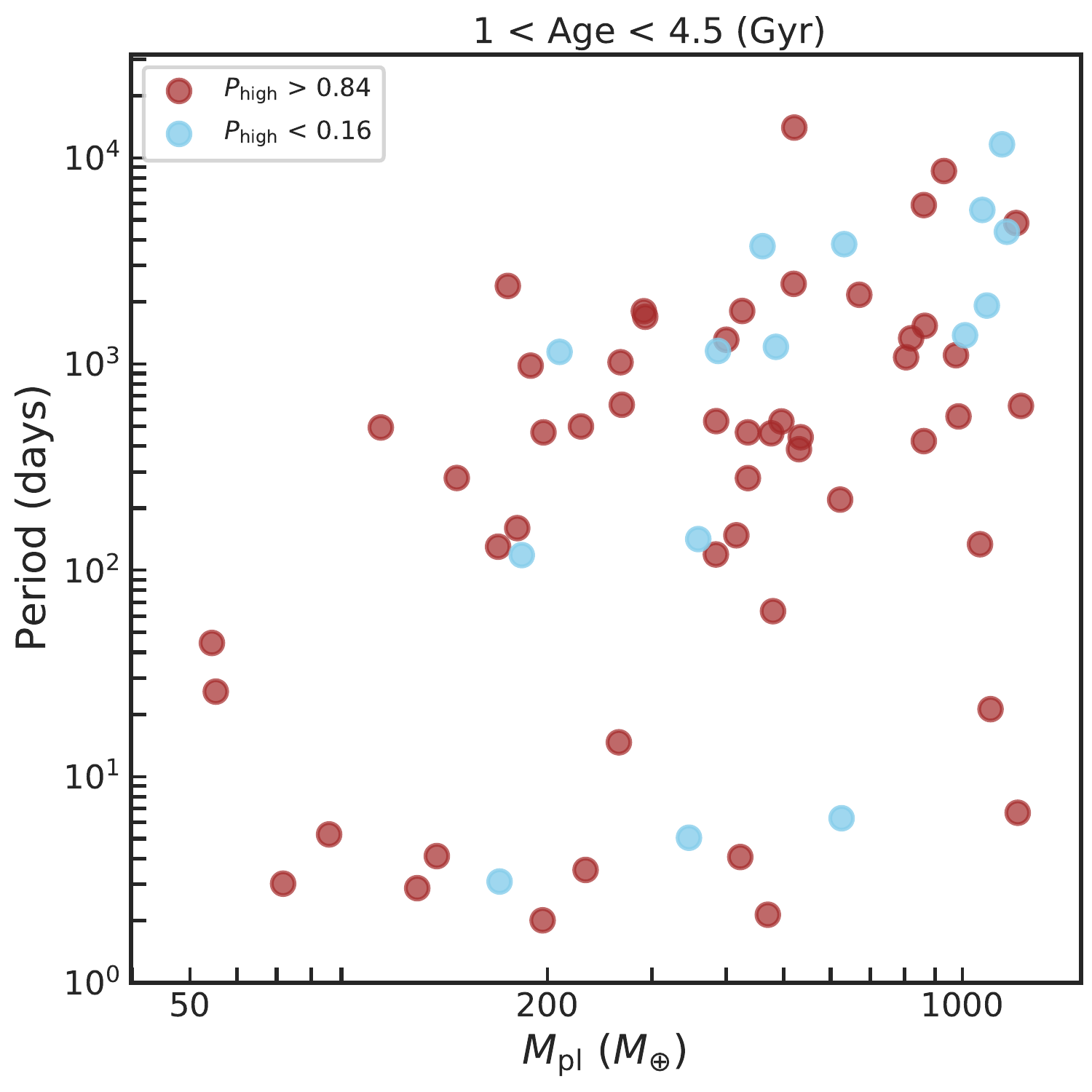}
 \includegraphics[angle=0,width=0.45\linewidth]{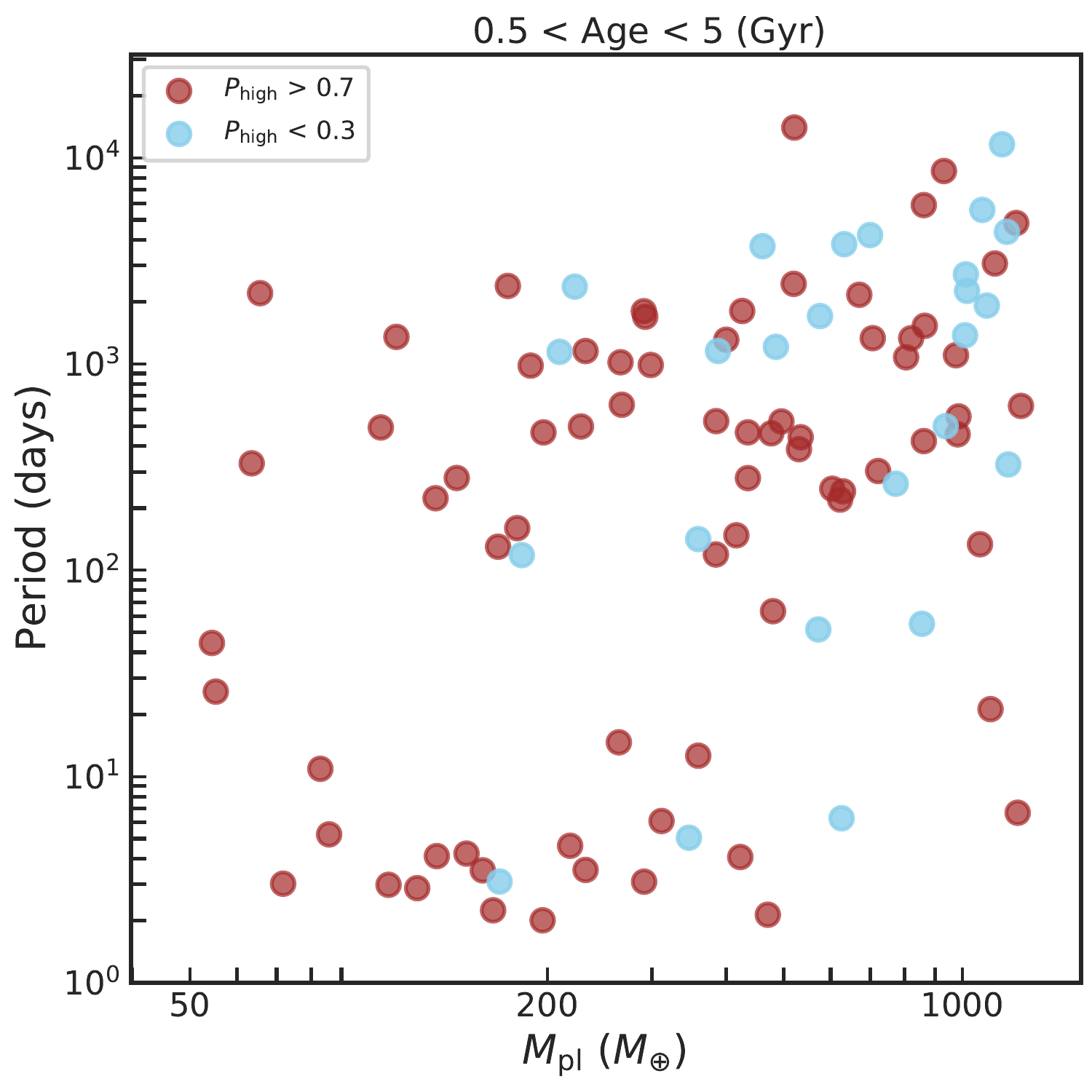}
\end{tabular}
\end{center}
\vspace{-0.5cm}
\caption{Distribution of the RV detected planets on the Period-Mass diagram orbiting around FGK dwarf stars with homogeneously derived stellar parameters in SWEET-Cat. The left panel is for the FGK$P^{\mathrm{low,high}}$  sample and the right panel is for the planets in the expended sample.}
\label{fig:P-M_diagram_sweetcat}
\end{figure*}

The aforementioned constraints lead us to a sample of 178 FGK dwarf stars hosting 214 RV detected giant planets\footnote{Note that due to the constraint on the homogeneity of the stellar parameters only 10 stars (i.e. 5\%) have been excluded at this stage.}. For all these stars we homogeneously determined the isochrone ages using the PARAM v1.3 web interface\footnote{http://stev.oapd.inaf.it/cgi-bin/param}. The detaials of the age determination and the results of their comparison with the ages used in W20 are presented in Sect.~\ref{sect:ages}. We then applied the final cut on age as suggested in W20 (stars with ages between 1 and 4.5 Gyr) to build our main sample, hereafter called FGK$P^{\mathrm{low,high}}$ sample. This sample consists of 44 $P^{\mathrm{high}}$\footnote{The stars with $P_{\mathrm{high}} >$ 0.84 are called $P^{\mathrm{high}}$, and the stars with $P_{\mathrm{high}} <$ 0.16 are called $P^{\mathrm{low}}$.} (52 planets) and 14 $P^{\mathrm{low}}$ (15 planets) stars. The distribution of these planets on the Period-Mass diagram is shown in the left panel of Fig.~\ref{fig:P-M_diagram_sweetcat}.

\subsection{Orbital periods of planets orbiting around $P^{\mathrm{low}}$ and $P^{\mathrm{high}}$ stars} \label{sect:W20_like_analysis}

In Fig.~\ref{fig:fgk_cumulative_plots_RV_main_sample} we compare the CDFs of different properties of planets and their host stars associated with over- and under-densities. The figure and the corresponding $P_{\mathrm{AD}}$ values suggest that the planets orbiting these two groups of stars do not have significantly different distributions of the orbital periods. The figure also shows that  the host star show significantly different distributions of $T_{\mathrm{eff}}$ and ages, the $P^{\mathrm{high}}$ stars being hotter and younger than their  $P^{\mathrm{low}}$ counterparts. In particular, only 3 out of the 15 planets orbiting $P^{\mathrm{low}}$ stars are younger than 3 Gyr. The number of planets (age $<$ 3 Gyr) orbiting young $P^{\mathrm{high}}$ stars is 32, which makes about 60\% (32 out of 52) of the whole sample.

\hspace{1cm}

Although in the aforementioned analysis the AD test does not rejects the null hypothesis that overall distributions of periods of planets orbiting  $P^{\mathrm{high}}$ and $P^{\mathrm{low}}$ stars come from the same parent distribution, Fig.~\ref{fig:P-M_diagram_sweetcat} visually suggest an overabundance of short period planets (periods shorter than about 10 to 30 days) around $P^{\mathrm{high}}$ stars when compared to their $P^{\mathrm{low}}$ counterparts.The fraction of short period (period $<$ 30 days\footnote{By adopting an upper limit of 0.2 AU for the semi-major axis of HJs, \citet{Winter-20} limited their sample to planets with orbital periods shorter than about 30 days.}) planets orbiting $P^{\mathrm{high}}$ stars is 23.1$^{\mathrm{+6.8}}_{\mathrm{-4.8}}$\% (12 out of 52). This number, being slightly larger, however, statistically speaking is not different from the one for the $P^{\mathrm{low}}$ sample,: 20.0$^{\mathrm{+13.7}}_{\mathrm{-6.6}}$\% (3 out of 15). The difference remains not significant if one considers more commonly used period limit of 10 days for HJs \citep[e.g.][]{Wang-15}: 17.3$^{\mathrm{+6.5}}_{\mathrm{-4.0}}$\% (9 out of 52) and 20.0$^{\mathrm{+13.7}}_{\mathrm{-6.6}}$\% (3 out of 15) for the HJs orbiting around $P^{\mathrm{high}}$ and $P^{\mathrm{low}}$ stars, respectively.

Unfortunately, by applying the cut on age and selecting only RV detected planets, we significantly reduced the size of the sample, especially the number of  $P^{\mathrm{low}}$ stars. The reduced sample size has a direct impact on the errors of the estimated HJs fractions and might be responsible for the  insignificance of the aforementioned differences. Below we try to expand the sample by increasing the range in stellar ages and relaxing the $P_{\mathrm{high}}$ threshold.

\begin{figure}
\begin{center}
\includegraphics[angle=0,width=1\linewidth]{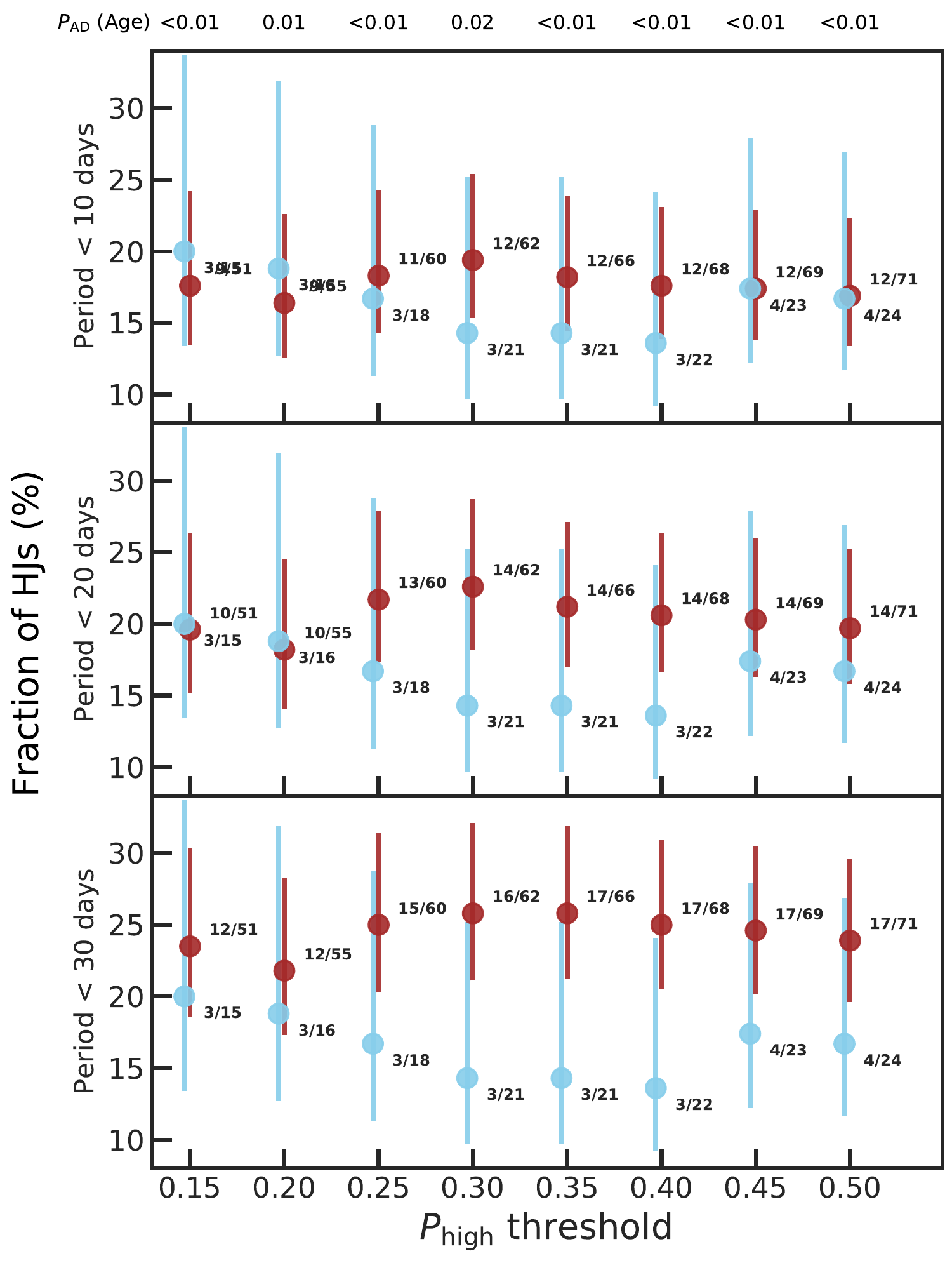}
\end{center}
\vspace{-0.5cm}
\caption{Fraction of HJs orbiting `cluster' (brown) and `field' (skyblue) stars  as a function of $P_{\mathrm{high}}$ threshold which is used to separate the aforementioned two groups. The numbers next to the HJs fraction correspond to the numbers of HJs and the total number of planets that are used to compute these fractions. The upper limit for the HJs orbital period is set at 10 (top), 20 (middle), and 30 (bottom) days. The $P_{\mathrm{AD}}$ values on top of the panels show the results of AD test comparing the distributions of the ages of the `cluster' and `field' stars. The errorbars represent 68.3\% (1-$\sigma$) interval.}
\label{fig:HJ_frequency_Phigh_threshold_main_sample}
\end{figure}

In Fig.~\ref{fig:HJ_frequency_Phigh_threshold_main_sample} we show fraction of HJs orbiting stars associated to over- and under-densities as a function of the $P_{\mathrm{high}}$ threshold which is used to separate the stars into this two categories. In the figure we considered 10, 20, and 30 days as the upper limit for the orbital periods of HJs. The figure shows that the maximum difference  of the fraction of HJs orbiting around `cluster' and `field' stars is observed at the $P_{\mathrm{high}}$ threshold of 0.3. This difference is however, is not significant at even one-$\sigma$ level. The figure also shows that, as for the main sample, the two groups of stars have significantly different distribution of ages.In the following tests we will adopt the 0.3 value as the $P_{\mathrm{high}}$ threshold, which allows both to increase the sample size and decrease the contamination of the samples by excluding the stars with intermediate $P_{\mathrm{high}}$ probabilities.

To further increase the sample size, we reduced the lower age limit from 1 Gyr to 0.5 Gyr. Although individual planets or planetary systems can show instabilities at timescales of a few Gyrs \citep[in fact, one of the phenomena responsible for the instability on long timescales is the fly-by encounters that can occur in dense stellar environments,][]{Davies-14}, usually the orbits of massive planets become stable at less than about 100 Myr \citep[e.g.][]{Raymond-09, Davies-14, Sotiriadis-17, Bitsch-20}. As discussed in W20 \citep[also see][]{Kruijssen-20}, going beyond 5 Gyr leads to a strong contamination of `field' sample by former over-density stars and should be avoided. In general, the younger the stars the easier and more reliable its association to over- or under-density stellar environments.

\begin{figure}
\begin{center}
\includegraphics[angle=0,width=1\linewidth]{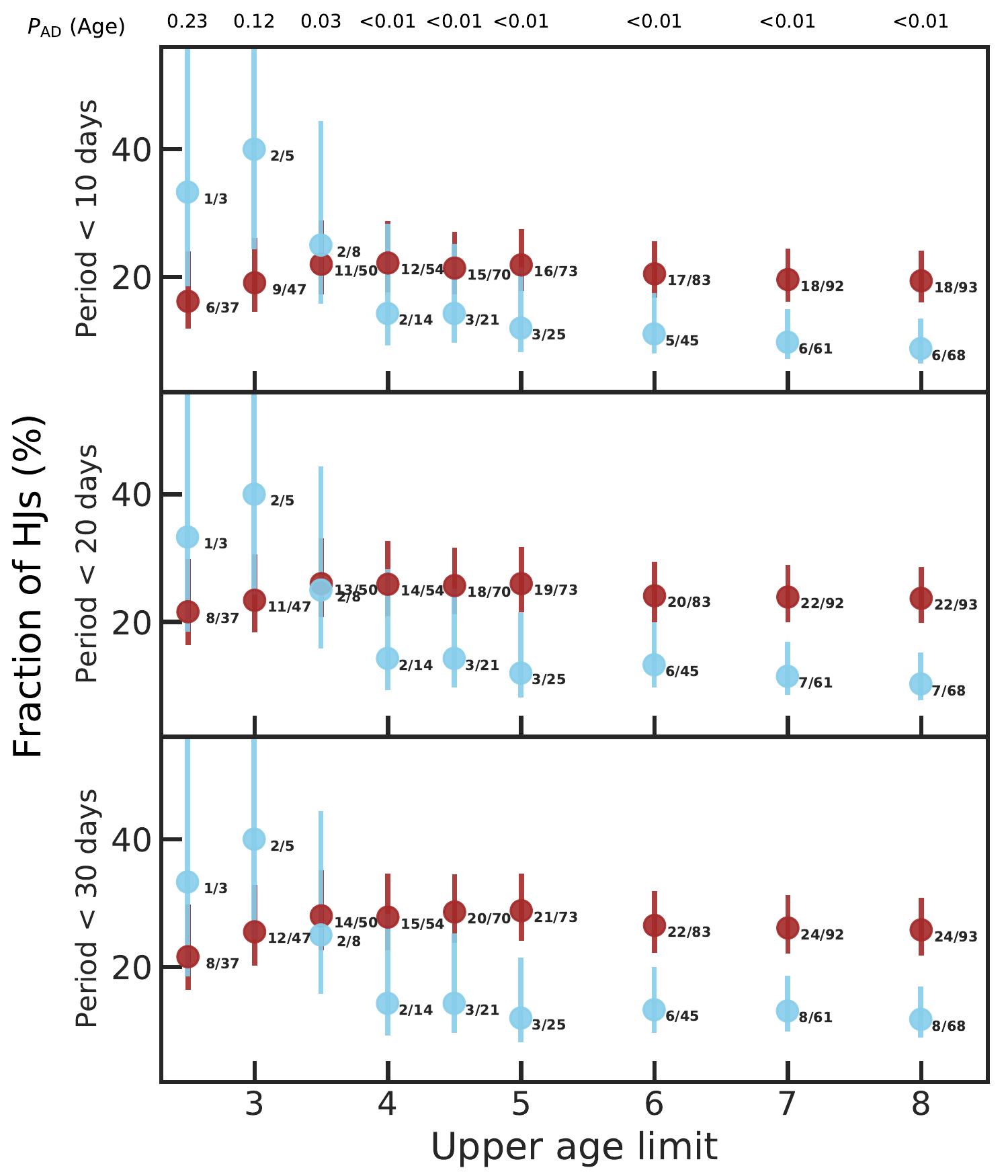}
\end{center}
\vspace{-0.5cm}
\caption{Fraction of HJs orbiting `cluster' (brown) and `field' (skyblue) stars  as a function of upper age  limit. The lower age limit is set to 0.5 Gyr and the $P_{\mathrm{high}}$ threshold is set to 0.3. The symbols and numbers are the same as in Fig.~\ref{fig:HJ_frequency_Phigh_threshold_main_sample}.}
\label{fig:HJ_frequency_age_threshold}
\end{figure}

Fig.~\ref{fig:HJ_frequency_age_threshold} shows the fraction of HJs orbiting stars associated to over- and under-densities as a function of upper limit of stellar ages. The figure shows that the HJs fractions are statistically speaking similar up to upper age limit of 5 Gyr. Moreover, the difference in the HJs fraction with age increases mostly because of the decrease of the fraction of HJs orbiting `field' stars. This is somehow counter-intuitive, because as it was mentioned earlier, older `field' samples are more contaminated by former `cluster' stars for which the fraction of HJs becomes higher. Up to ages of 3 Gyr, the fraction of HJs orbiting around `field' stars is even higher than that for the `cluster' stars. However, it is important to note that the number of  $P^{\mathrm{low}}$ stars with ages below 3 Gyr is very small. Fig.~\ref{fig:HJ_frequency_age_threshold} also shows that when going beyond the 5 Gyr limit, the HJ fraction slightly decreases for both $P^{\mathrm{low}}$ and $P^{\mathrm{high}}$ samples. This is because on average the hosts of HJs are slightly younger than the hosts of their longer period counterparts. This result is similar to the one of \citet{Hamer-19} where the authors concluded that tidal interactions cause HJs to inspiral on a timescale shorter than the main sequence lifetime of the stars.

Considering stars with ages between 0.5 and 5 Gyr, and the $P_{\mathrm{high}}$ threshold of 0.3 we construct an extended sample consisting of 73 planets orbiting $P^{\mathrm{high}}$ stars and 25 planets orbiting $P^{\mathrm{low}}$ stars. The distributions of these planets in the Period-Mass diagram is shown in the right panel of Fig.~\ref{fig:P-M_diagram_sweetcat}. The difference in the HJ fractions between the two groups is largest when considering an upper orbital period limit of 30 days for HJs. This difference (28.8$^{\mathrm{+5.8}}_{\mathrm{-4.7}}$\% (21 out of 73) and 12.0$^{\mathrm{+9.5}}_{\mathrm{-3.8}}$\% (3 out of 25) is significant at about 80\% level, which would correspond to $\sim$ 1.3-$\sigma$ for a Gaussian distribution. However, it is important to stress again the statistically significant difference in age as inferred from the p-values of the AD test (see Fig.~\ref{fig:fgk_cumulative_plots_RV_extended_sample}).

For the extended sample Fig.~\ref{fig:fgk_cumulative_plots_RV_extended_sample} suggest a statistically significant difference for the period distributions of planets orbiting stars in over- and under-densities. However, the two sub-samples show statistically different distributions in planetary mass, $T_{\mathrm{eff}}$, and stellar age. Restricting the sample (25 and 22 planets orbiting around $P^{\mathrm{high}}$ and $P^{\mathrm{low}}$ stars, respectively) to stars with ages between 2.5 and 5 Gyr and to planets with masses $>$ 150 M$_{\oplus}$ allows to vanish the differences in planetary mass, $T_{\mathrm{eff}}$, and stellar age. This restriction also dilutes the difference in the orbital period distributions (see Fig.~\ref{fig:fgk_cumulative_plots_RV_restricted_sample}).

\section{Properties of stars hosting short- versus long-period planets in over- and under-density environments}
\label{sect:W20_reversed_analysis}

The analysis of the previous section did not reveal an unambiguous relation between stellar clustering and orbital architecture of exoplanets. In this section, we separate the $P^{\mathrm{high}}$ and $P^{\mathrm{low}}$ samples to study the impact of physical properties of the host stars on the orbital properties of giant planets. In this way we eliminate the impact (if any) of stellar clustering on the architecture of planets.

\citet{Adibekyan-13} showed that most of the massive planets orbiting low-metallicity stars ([Fe/H] $<$ -0.1 dex) have orbital periods longer than about 100 days \citep[see also][]{Sozzetti-04, Maldonado-12}. The authors suggested that planets in a metal-poor disk are forming further out and/or undergoing less migration as they take longer to form. Recently, \citet{Osborn-20} studied the metallicity distribution of HJs and found that although they preferentially orbit metal-rich stars, the average metallicity of their hosts is not higher than that of stars hosting cold Jupiters. The authors concluded that hot and cold Jupiters are formed in a similar process, but they have different migration histories. In complement to these results, \citet{Dawson-13} and \citet{Buchhave-18} showed that giant planets orbiting metal-rich stars show signatures of dynamical interaction. \citet{Buchhave-18}, in particular, showed that HJs and cold eccentric Jupiters are preferentially orbiting metal-rich stars, while cold-Jupiters with circular orbits are mostly observed around solar-metallicity stars.

\begin{figure*}
\begin{center}
\includegraphics[angle=0,width=0.7\linewidth]{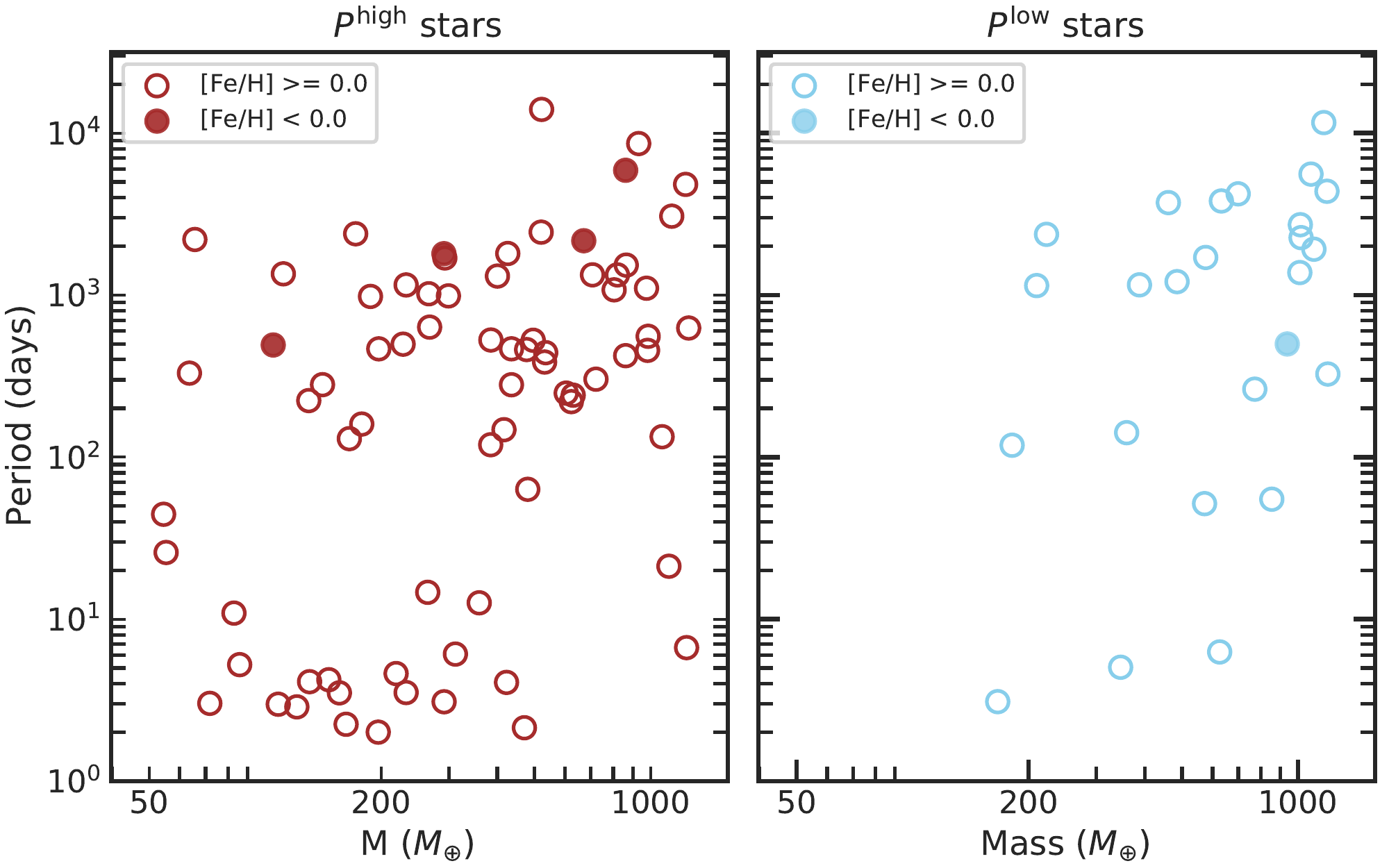}
\end{center}
\vspace{-0.5cm}
\caption{Period-Mass diagram of planets around FGK $P^{\mathrm{high}}$ (left panel) and $P^{\mathrm{low}}$ (right panel) dwarf stars with homogeneously derived stellar parameters in SWEET-Cat. Only stars with ages between 1 and 4.5 Gyr are shown. Planets orbiting metal-poor and metal-rich stars are shown in open and filled circles, respectively.}
\label{fig:P-M_diagram_sweetcat_inverse_young}
\end{figure*}

On the left and right panels of Fig.~\ref{fig:P-M_diagram_sweetcat_inverse_young} we separately show the distribution of planets orbiting the $P^{\mathrm{high}}$ and $P^{\mathrm{low}}$ stars of the extended FGK$P^{\mathrm{low,high}}$ sample. Due to the age constraints (young stars are on average metallic), the number of planets orbiting metal-poor stars (Fe/H] $<$ 0.0 dex) in this sample is very small. However, as was found in \citet{Adibekyan-13}, they all have orbital periods longer than 100 days. 

Because the sample of $P^{\mathrm{low}}$ stars is very small, we will next focus only on the planets orbiting $P^{\mathrm{high}}$ stars. This sample consists of 21 planets with periods shorter than 30 days and 52 planets with longer periods. Fig.~\ref{fig:fgk_young_cumulative_plots_Phigh_short_long} shows the CDFs of these short- and long-periods planets and their host stars. As indicated by the $P_{\mathrm{AD}}$ values, the two groups are significantly different only in the distribution of the planetary masses, the HJs having on average lower-masses. If the upper period limit is reduced to 10 or 20 days, the results remain practically the same. The fact that the age distribution of the short and long period planets are similar might indicate that tidal inspiral is not significantly depleting the short period planets in this age range. However, a dedicated analysis on a larger sample is required to make a firm conclusion.

Unfortunately, perhaps due to the small size of the sample and/or due to the complexity of the problem, it is difficult to firmly conclude which parameter(s) internal for the star-planet system is(are) responsible for the period distribution of exoplanets orbiting stars formed in similar stellar environment.

\section{Summary and conclusion} \label{sect:summary}

Very recently \citet{Winter-20} performed a tremendous work by assigning a large sample of exoplanet host stars to low- ($P^{\mathrm{low}}$) or high-density ($P^{\mathrm{high}}$) stellar environments. The authors then used this sample to conclude that planets orbiting stars in high-density environments have significantly shorter periods and smaller semi-major axis than their  counterparts orbiting `field' (low-density) stars. They also found that most of the hot Jupiters are orbiting around `cluster' stars. These findings, if confirmed, may have very important implications for our understanding of planet formation and evolution.

In this manuscript we constructed a sample of FGK dwarf stars with only RV detected HJs for which homogeneously determined stellar parameters are available in the SWEET-Cat catalog \citep{Santos-13}. Additionally, we made a further constraints on the upper mass of planets at 4~$M_{\mathrm{jup}}$ since the origin of the super-massive planets might be different \citep[e.g.][]{Santos-17, Adibekyan-19}. For this sample of stars we homogeneously determined isochrone ages. 

In this small but significantly less biased sample of stars with ages between 1 and 4.5 Gyr (52 planets orbiting $P^{\mathrm{high}}$ stars and 15 planets orbiting $P^{\mathrm{low}}$ stars), we found no significant difference in the period distribution of planets orbiting $P^{\mathrm{high}}$ and  $P^{\mathrm{low}}$ stars. We then constructed an extended sample by slightly relaxing the constrains on age and the $P_{\mathrm{high}}$ threshold. In this sample, consisting of 73 planets orbiting around 'cluster' stars and 25 planets orbiting around 'filed' stars, we found a statistically significant difference for the period distributions of planets orbiting around these two populations of stars. However, the `field' and `cluster' stars also showed a significant difference in the stellar age. When controlling for the host star properties, the differences in orbital periods of planets orbiting around stars associated with the over- and under-densities diminishes.  Thus, it is not possible to conclude whether the planetary architecture is related to age, environment, or both.

Next we focused only on a sub-sample of planets orbiting $P^{\mathrm{high}}$ stars with the aim of understanding the mechanism responsible for shaping their planetary orbits in similar environments. We could not identify a parameter that unambiguously can be responsible for the orbital architecture of these planets.

It is important to note that although our analysis does not suggest that the stellar clustering is the key parameter shaping the orbits of planets, it still can play a role, especially given some observational \citep[e.g.][]{Brucalassi-16} and theoretical \citep[e.g.][]{Shara-16, Wang-20} support of this hypothesis. The full picture of planet survival in dense stellar environments is not simple and depends on many external and internal to star-planets factors \citep[e.g.][and references therein]{Stock-20}. Increased number of planet hosts in clusters and in over-density environments will help to build large and unbiased samples which will then shed a light on this issue.

\begin{acknowledgements}
 We thank the anonymous referee for the very constructive comments and suggestions which helped us to substantially improve the quality of the work amd presentations of the results. 
 This work was supported by FCT - Funda\c{c}\~ao para a Ci\^encia e Tecnologia (FCT) through national funds and by FEDER through COMPETE2020 - Programa Operacional Competitividade e Internacionaliza\c{c}\~ao by these grants: UID/FIS/04434/2019; UIDB/04434/2020; UIDP/04434/2020; PTDC/FIS-AST/32113/2017 \& POCI-01-0145-FEDER-032113; PTDC/FIS-AST/28953/2017 \& POCI-01-0145-FEDER-028953. V.A., E.D.M, N.C.S., and S.G.S. also acknowledge the support from FCT through Investigador FCT contracts nr.  IF/00650/2015/CP1273/CT0001, IF/00849/2015/CP1273/CT0003, IF/00169/2012/CP0150/CT0002,   and IF/00028/2014/CP1215/CT0002, respectively, and POPH/FSE (EC) by FEDER funding through the program ``Programa Operacional de Factores de Competitividade - COMPETE''. O.D.S.D. and  J.P.F. are supported in the form of work contracts (DL 57/2016/CP1364/CT0004 and DL57/2016/CP1364/CT0005, respectively) funded by FCT. T.C.~is supported by Funda\c c\~ao para a Ci\^encia e a Tecnologia (FCT) in the form of a work contract (CEECIND/00476/2018).

This research has made use of the NASA Exoplanet Archive, which is operated by the California Institute of Technology, under contract with the National Aeronautics and Space Administration under the Exoplanet Exploration Program.

In this work we used the Python language and several scientific packages: Numpy\cite{van_der_Walt-11}, Scipy 
\cite{Virtanen-20}, Pandas \cite{mckinney-proc-scipy-2010}, Astropy \cite{Astropy-18}, and Matplotlib \cite{Hunter-07}.
\end{acknowledgements}

\bibliography{references.bib}

\clearpage

\begin{appendix}

\section{Importance of homogeneity of stellar parameters} \label{sect:parameters}

When performing a statistical analysis of the properties of stars with and/or without planets, it is important to use parameters as homogeneously derived as possible \citep[e.g.][]{Adibekyan-19}. Exoplanet archives and catalogs usually consist of  heterogeneous compilation of stellar properties which might lead to significant discrepancies when compared with homogeneously derived parameters \citep[e.g.][]{Santos-13, Sousa-18}. The host star properties listed in NEA (used by W20) are compiled from different sources. Moreover, while the physical parameters of the RV detected planets are mostly derived from high-resolution spectra, such high-quality data do not necessarily exist for the transiting planet hosts. 

\begin{figure}
\begin{center}
\includegraphics[width=0.85\linewidth]{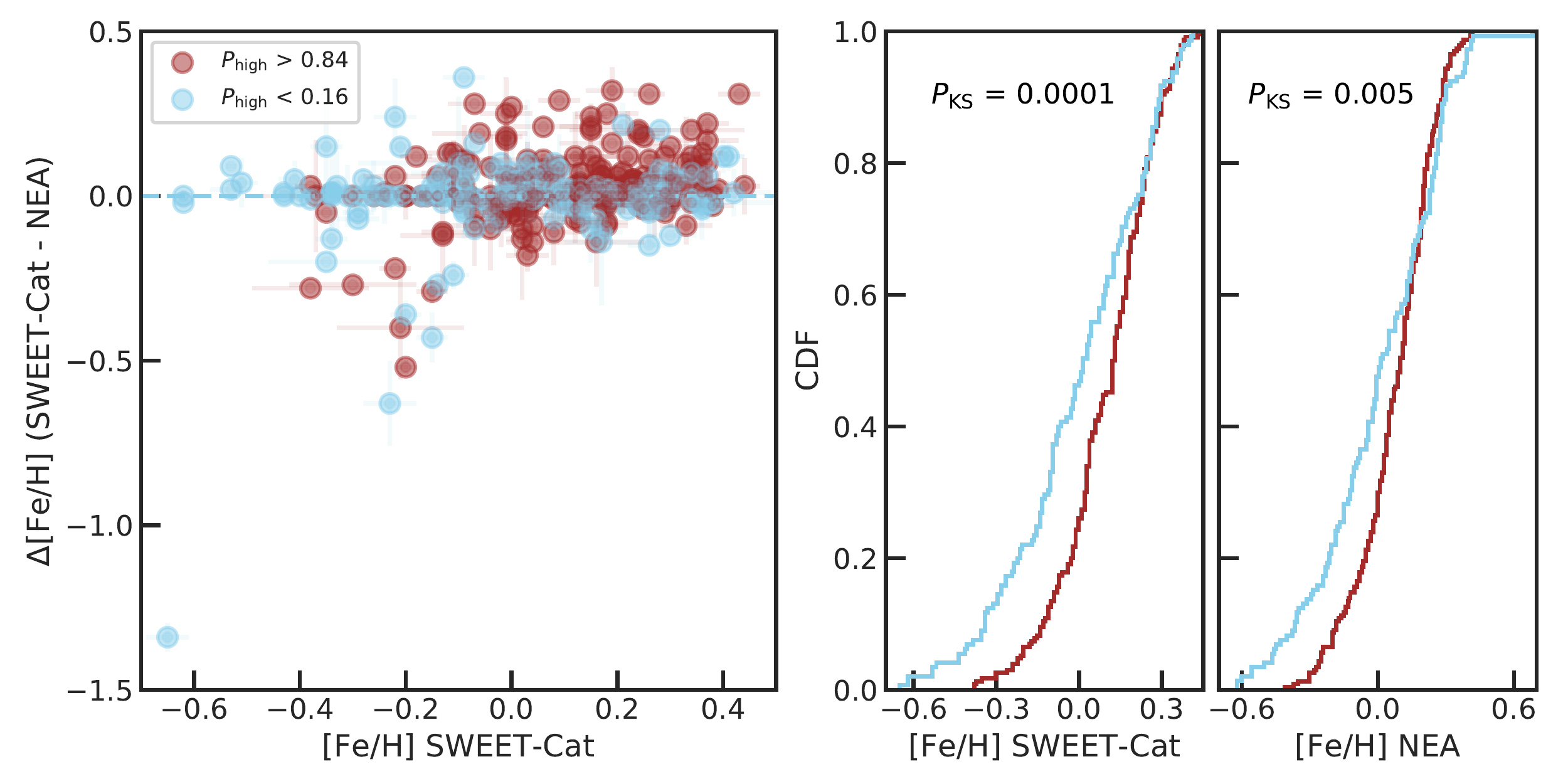}\\
\includegraphics[width=0.85\linewidth]{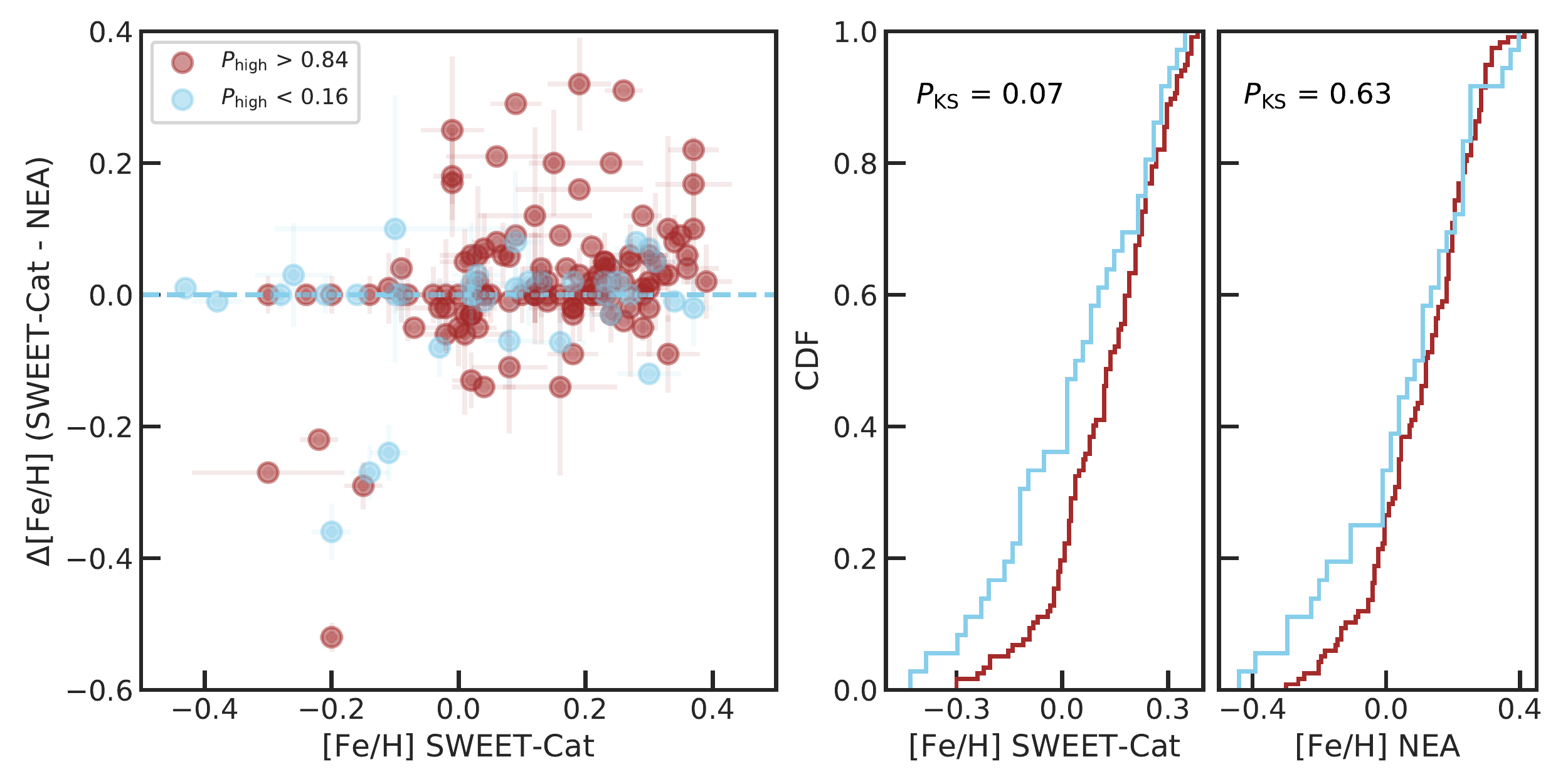}
\end{center}
\caption{Comparison of the stellar metallicities taken from the NASA Exoplanet Archive (NEA) and homogeneously derived in SWET-Cat. The top and bottom panels show the results for the full and main samples, respectively. The brown symbols and curves represent the data for the $P^{\mathrm{high}}$ stars and the skyblue symbols and curves for the $P^{\mathrm{low}}$ stars. The right panels show the CDFs of the metallicities for the $P^{\mathrm{low}}$ and $P^{\mathrm{high}}$ stars.}
\label{fig:metallicity}
\end{figure}

\begin{figure}
\begin{center}
\includegraphics[width=0.8\linewidth]{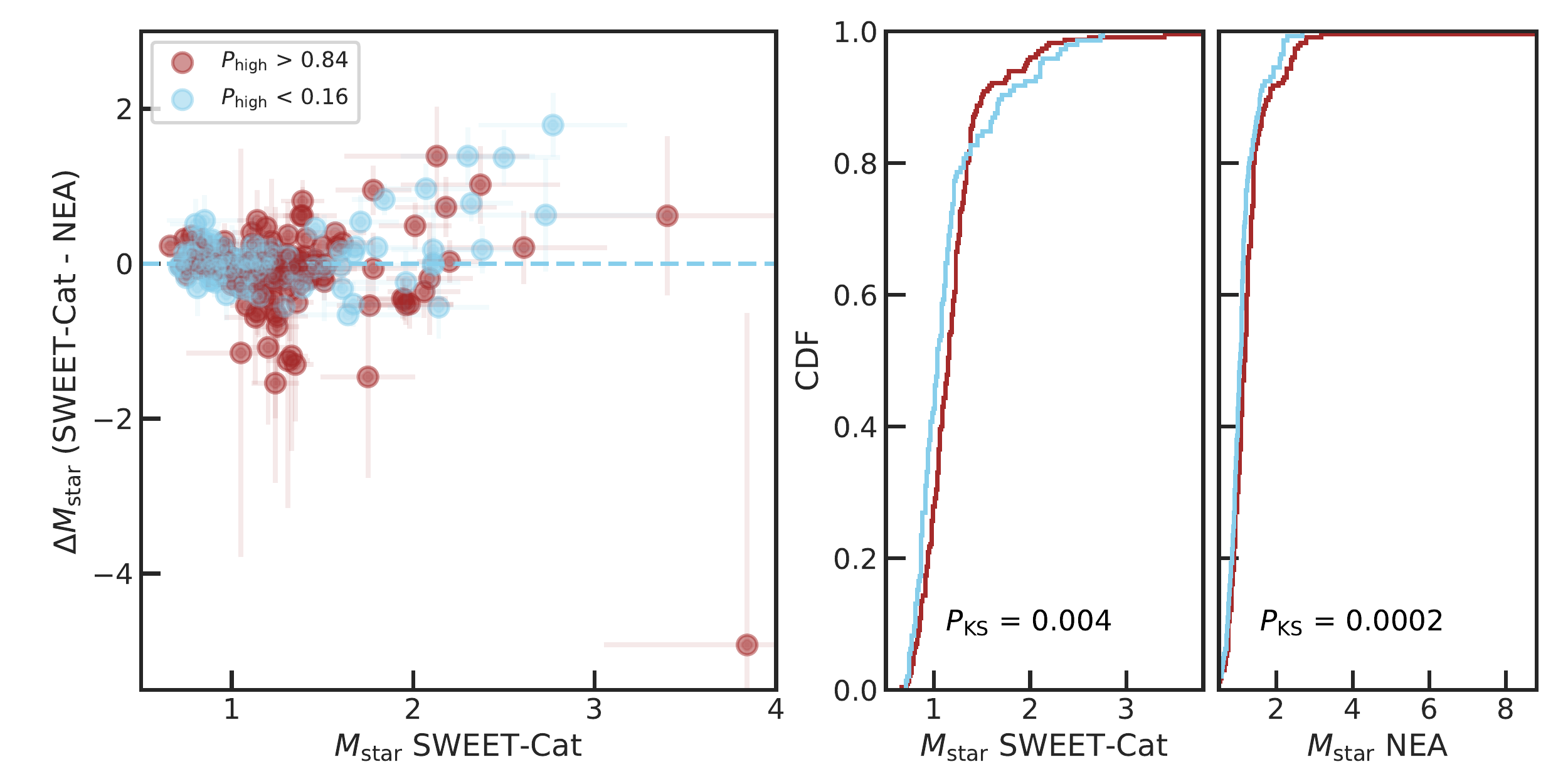}\\
\includegraphics[width=0.8\linewidth]{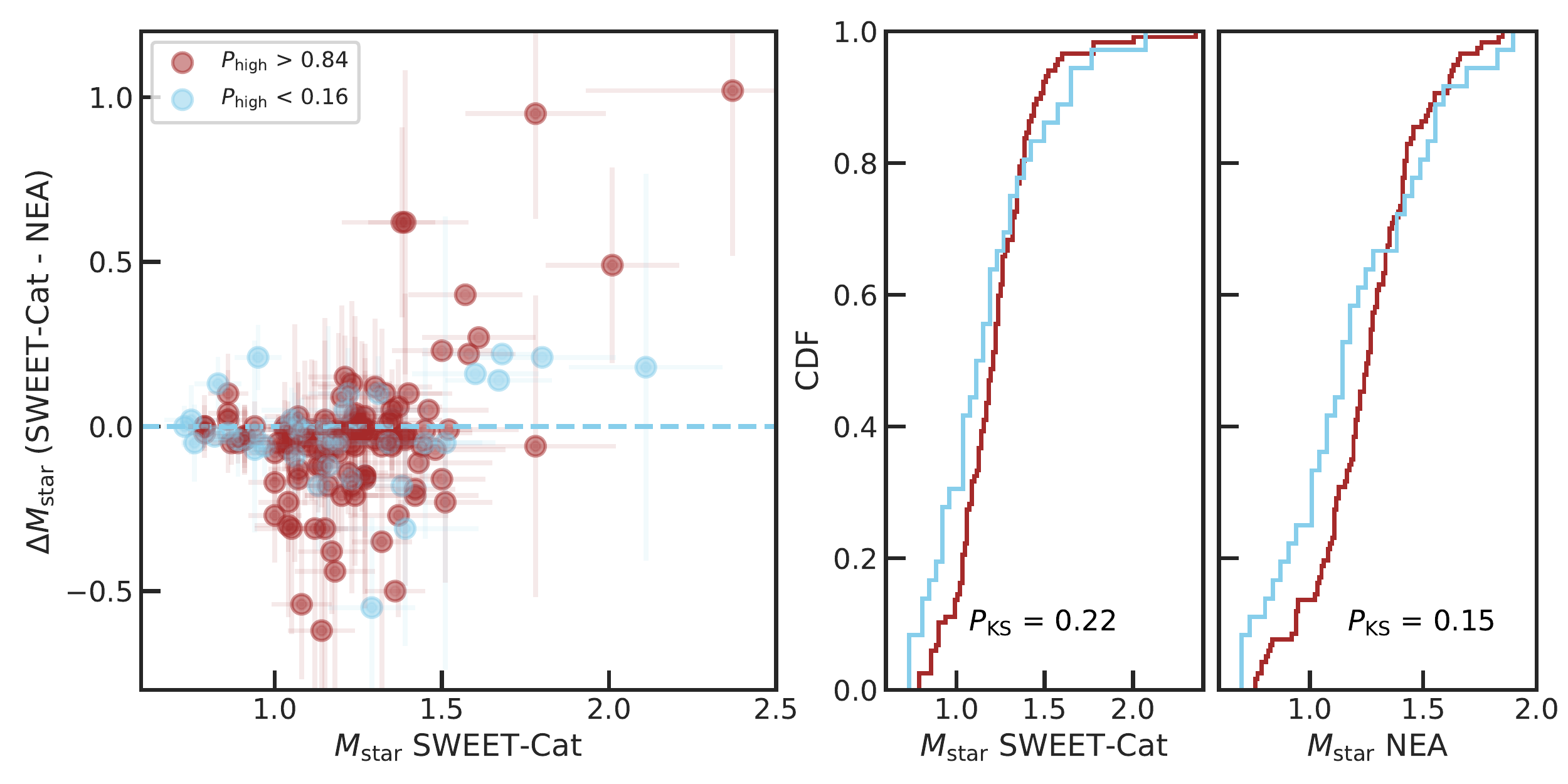}
\end{center}
\caption{The same as Fig.~\ref{fig:metallicity} but for stellar masses.}
\label{fig:mass}
\end{figure}

We cross-matched the full (1421 planets orbiting around 1058 stars) and the main (506 planets orbiting around 388 stars) samples of W20 with the SWEET-Cat catalog \citep{Santos-13, Sousa-18} which provides the stellar parameters of planet host stars. Although SWEET-Cat is one of the largest catalog (and the largest one for the RV detected planets) of planet host stars with homogeneously determined stellar parameters, unfortunately it contains stellar parameters only for 375 stars from the full sample and 153 stars from the main sample of W20. In Figs.~\ref{fig:metallicity} and \ref{fig:mass} we compare the stellar metallicities and masses presented in NEA and homogeneously derived in SWEET-Cat catalog \citep{Santos-13}. The mean difference and dispersion for metallicity is 0.01$\pm$0.12 dex and 0.01$\pm$0.10 dex for the full and  main samples, respectively. For the stellar masses, the mean difference and dispersion is 0.0$\pm$0.4 $M_{\odot}$ and 0.0$\pm$0.2 $M_{\odot}$ for the full and  main samples, respectively.

In the left panels of Figs.~\ref{fig:metallicity} and \ref{fig:mass} for the $P^{\mathrm{high}}$ and $P^{\mathrm{low}}$ stars we compare the CDFs of metallicity and masses as taken from NEA and SWEET-Cat. We then performed a KS test to evaluate the similarities of the distributions. The $P_{\mathrm{KS}}$ values for most of the cases are very similar. The exception is for the stellar metallicity for the main sample, where there is a significant difference of $P_{\mathrm{KS}}$ obtained for the NEA and SWEET-Cat values. Furthermore, if instead of the KS test the  Anderson-Darling (AD)\footnote{AD is similar to the KS test, but is more sensitive to the tails of distribution and has higher power for small samples \citep{Pettitt-76, Saculinggan-2013}.} test is performed to the aforementioned samples, the result would differ more dramatically ($P_{\mathrm{AD}}$ = 0.03 and $P_{\mathrm{AD}}$ $>$ 0.25 for the SWEET-Cat and NEA metallicities, respectively) suggesting that the metallicity (taken from SWEET-Cat) distributions of the $P^{\mathrm{high}}$ and $P^{\mathrm{low}}$ stars do not come from the same parent distribution.

\section{Stellar ages} \label{sect:ages}

For the sample of 178 FGK dwarf stars hosting 214 RV detected giant planets (see Sect.~\ref{sect:sweet-cat}) we derived the stellar ages from the PARAM v1.3 web interface\footnote{http://stev.oapd.inaf.it/cgi-bin/param} based on the Padova theoretical isochrones from \citet{Bressan-12} and with the use of a Bayesian estimation method \citep[][]{daSilva-06}. As input parameters for PARAM, we used the Gaia DR2 parallaxes \citep{Gaia-18}, V magnitudes extracted from Simbad\footnote{http://simbad.u-strasbg.fr/simbad/}, and spectroscopic $T_{\mathrm{eff}}$ and [Fe/H]. No correction for interstellar reddening was needed since all the stars are nearby objects. The ages  of all the stars is presented in a table at the CDS.

\begin{figure}
\begin{center}
\includegraphics[angle=0,width=0.8\linewidth]{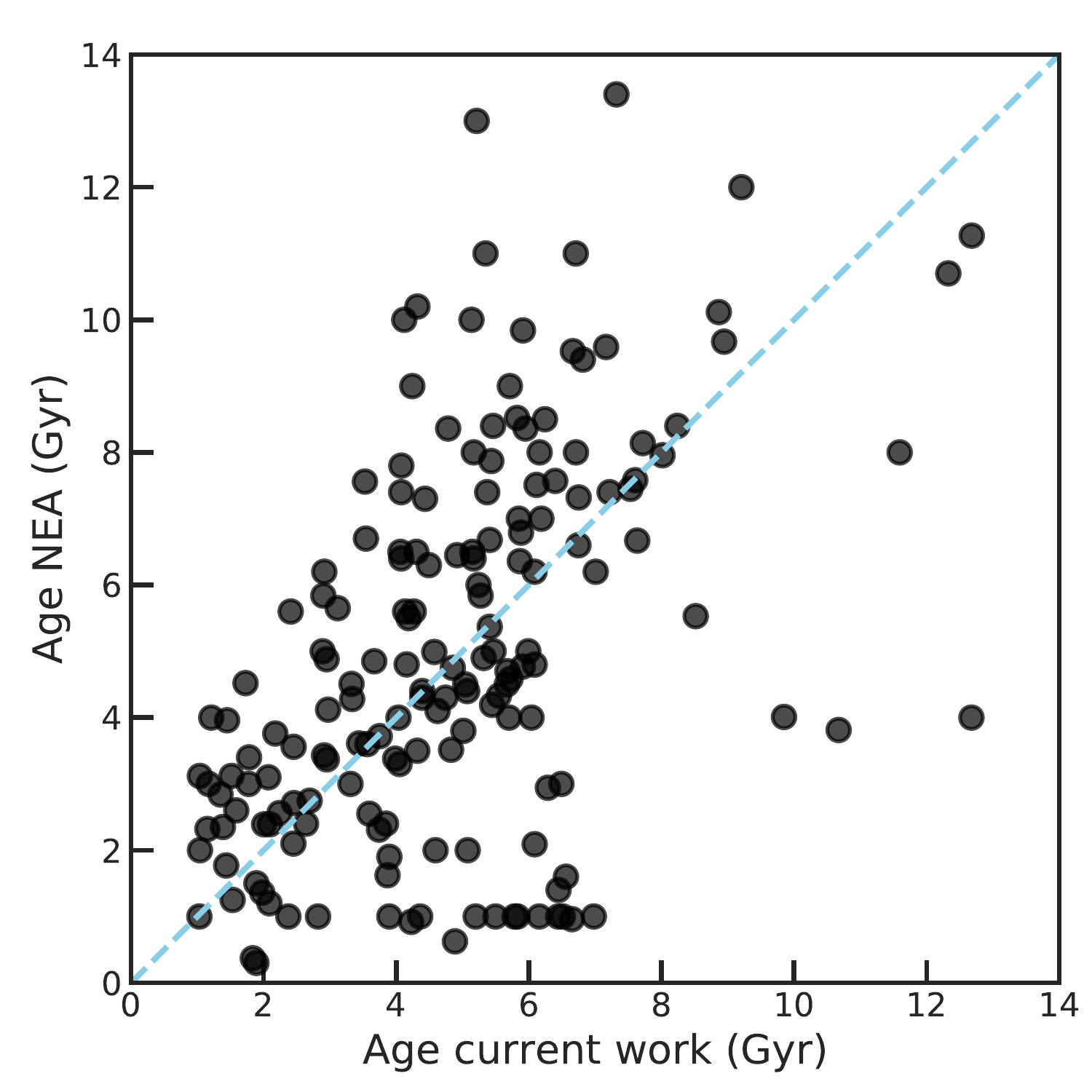}
\end{center}
\vspace{-0.5cm}
\caption{Comparison of the stellar ages taken from NEA and homogeneously derived in this work. The skyblue dashed line represents the identity line.}
\label{fig:age_comparision}
\end{figure}

In Fig.~\ref{fig:age_comparision} we compare the ages homogeniously derived in this work and those from NEA. While practically there is no offset (-0.1 Gyr) the dispersion is 2.7 Gyr. The figure also shows a group of 13 stars with NEA ages of exactly 1 Gyr. Eight of these stars, however, have isochrone ages (as derived in this work) greater than 5 Gyr and are excluded from the main sample. All these stars all cool ($T_{\mathrm{eff}} <$ 5350 K) and slightly evolved ($\log g \ < 4.4$ dex) indicating about their non-young ages.

\section{Supplementary figures}

\begin{figure*}
\begin{center}
\includegraphics[angle=0,width=0.8\linewidth]{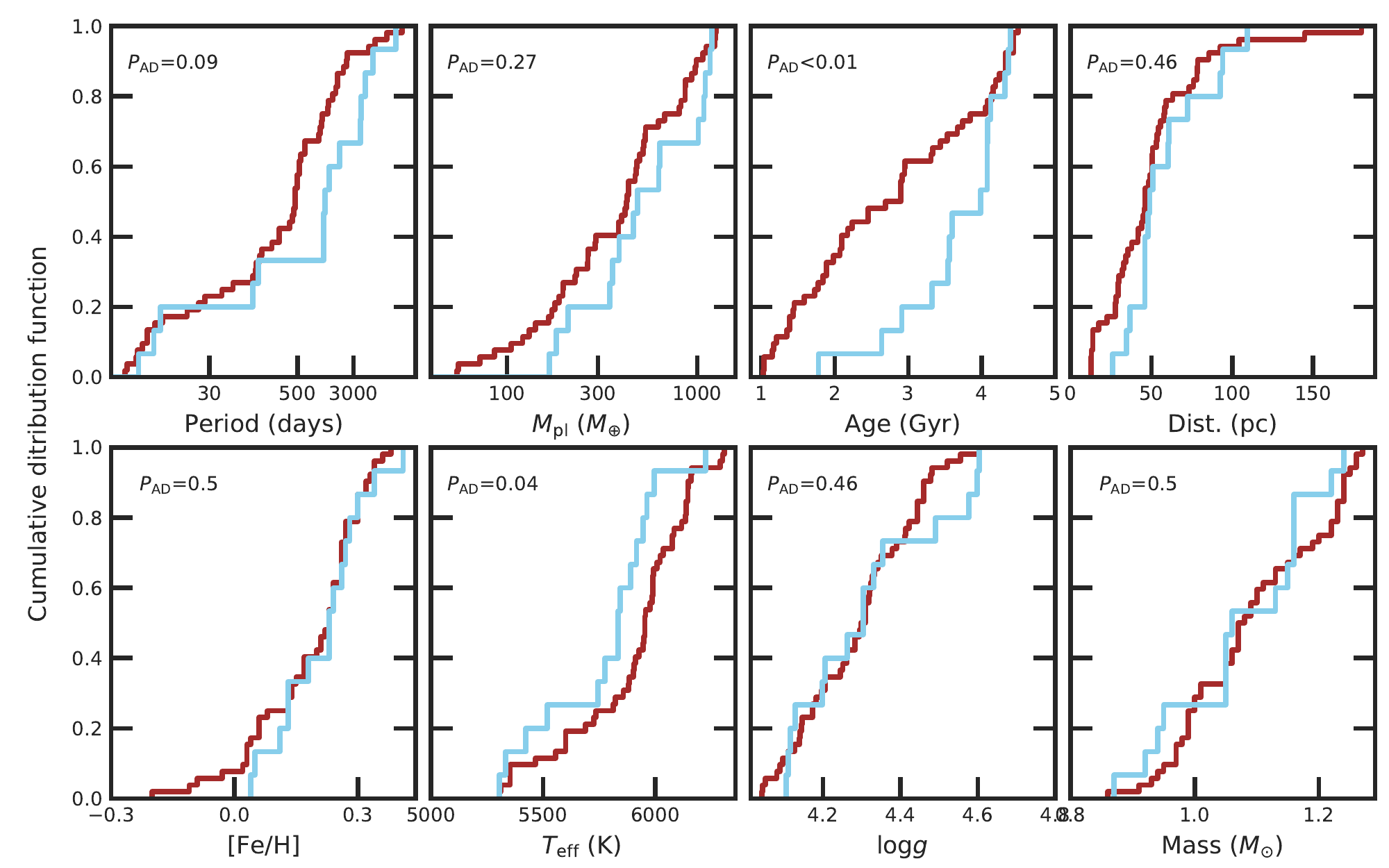}
\end{center}
\vspace{-0.5cm}
\caption{CDFs of different properties of planets and their host stars from the FGK$P^{\mathrm{low,high}}$ sample associated with over- (brown) and under-densities (skyblue). The $P_{\mathrm{AD}}$ values for each parameter is shown in the respective plot.}
\label{fig:fgk_cumulative_plots_RV_main_sample}
\end{figure*}

\begin{figure*}
\begin{center}
\includegraphics[angle=0,width=0.8\linewidth]{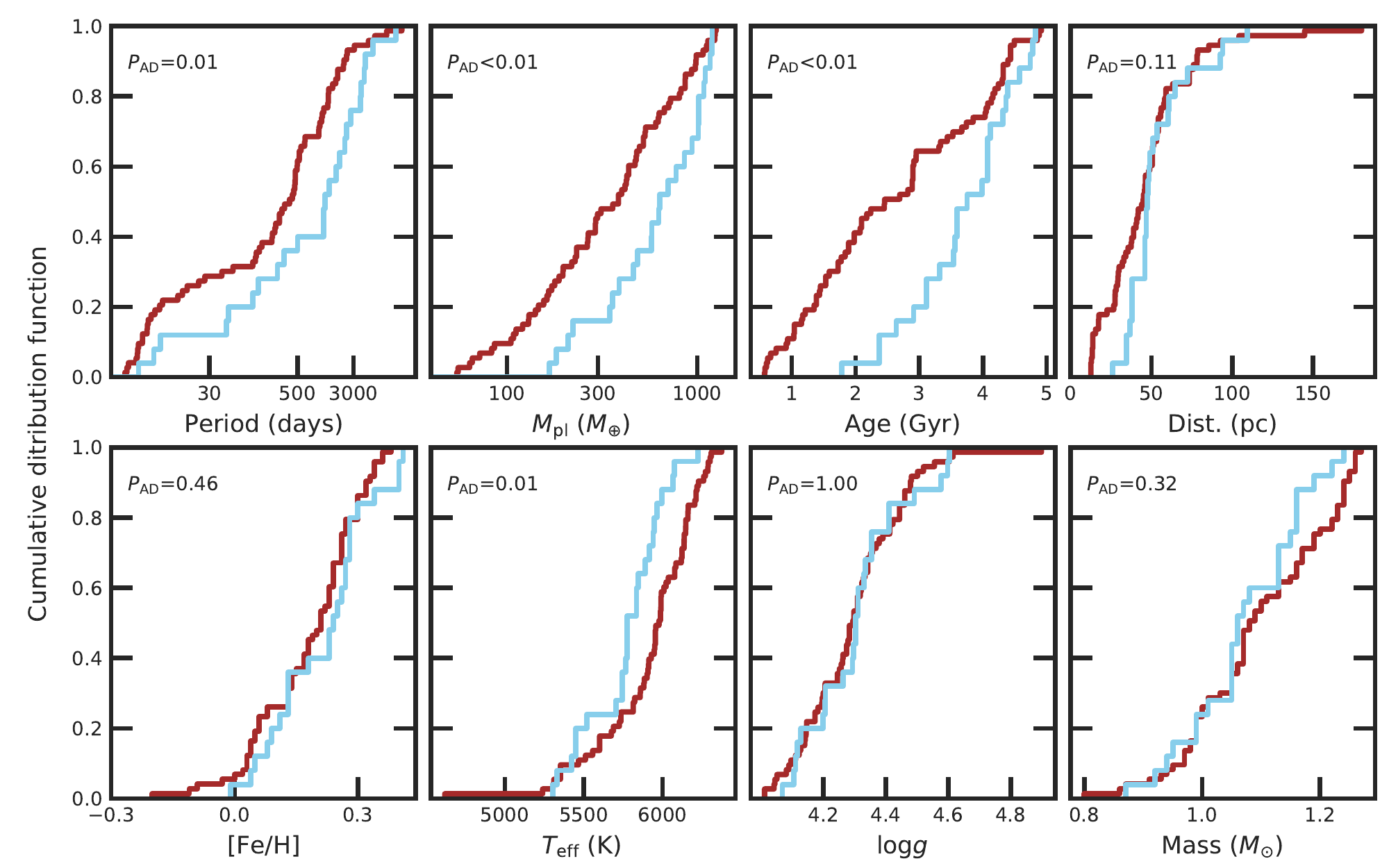}
\end{center}
\vspace{-0.5cm}
\caption{The sames as Fig.~\ref{fig:fgk_cumulative_plots_RV_main_sample} but for the extended sample: stars with ages between 0.5 and 5 Gyr, and the $P_{\mathrm{high}}$ threshold of 0.30.}
\label{fig:fgk_cumulative_plots_RV_extended_sample}
\end{figure*}

\begin{figure*}
\begin{center}
\includegraphics[angle=0,width=0.8\linewidth]{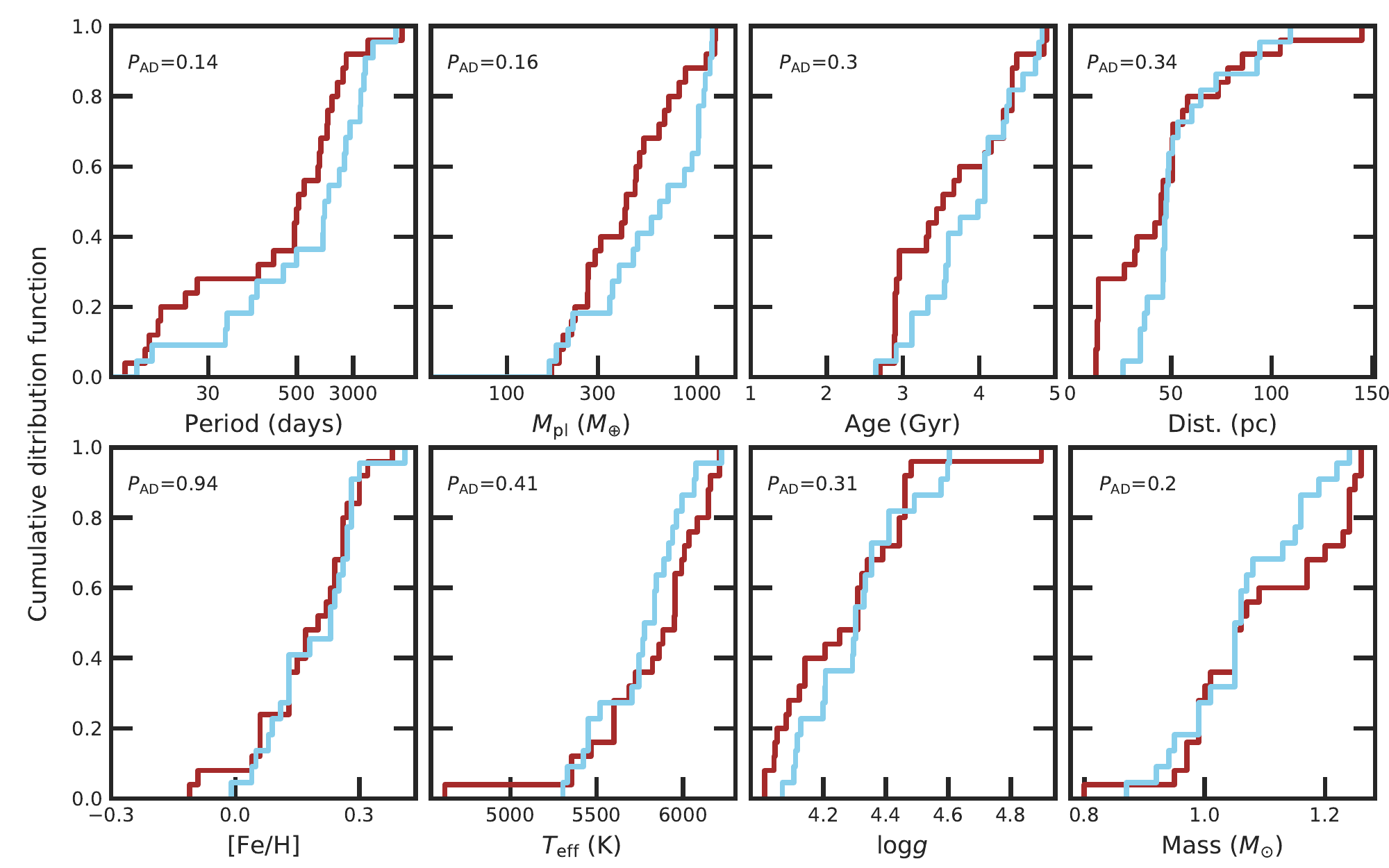}
\end{center}
\vspace{-0.5cm}
\caption{The sames as Fig.~\ref{fig:fgk_cumulative_plots_RV_main_sample} but for a sample of stars with ages between 2.5 and 5 Gyr, planets with masses greater than 150 M$_{\oplus}$, and the $P_{\mathrm{high}}$ threshold of 0.30.}
\label{fig:fgk_cumulative_plots_RV_restricted_sample}
\end{figure*}

\begin{figure*}
\begin{center}
\includegraphics[angle=0,width=0.9\linewidth]{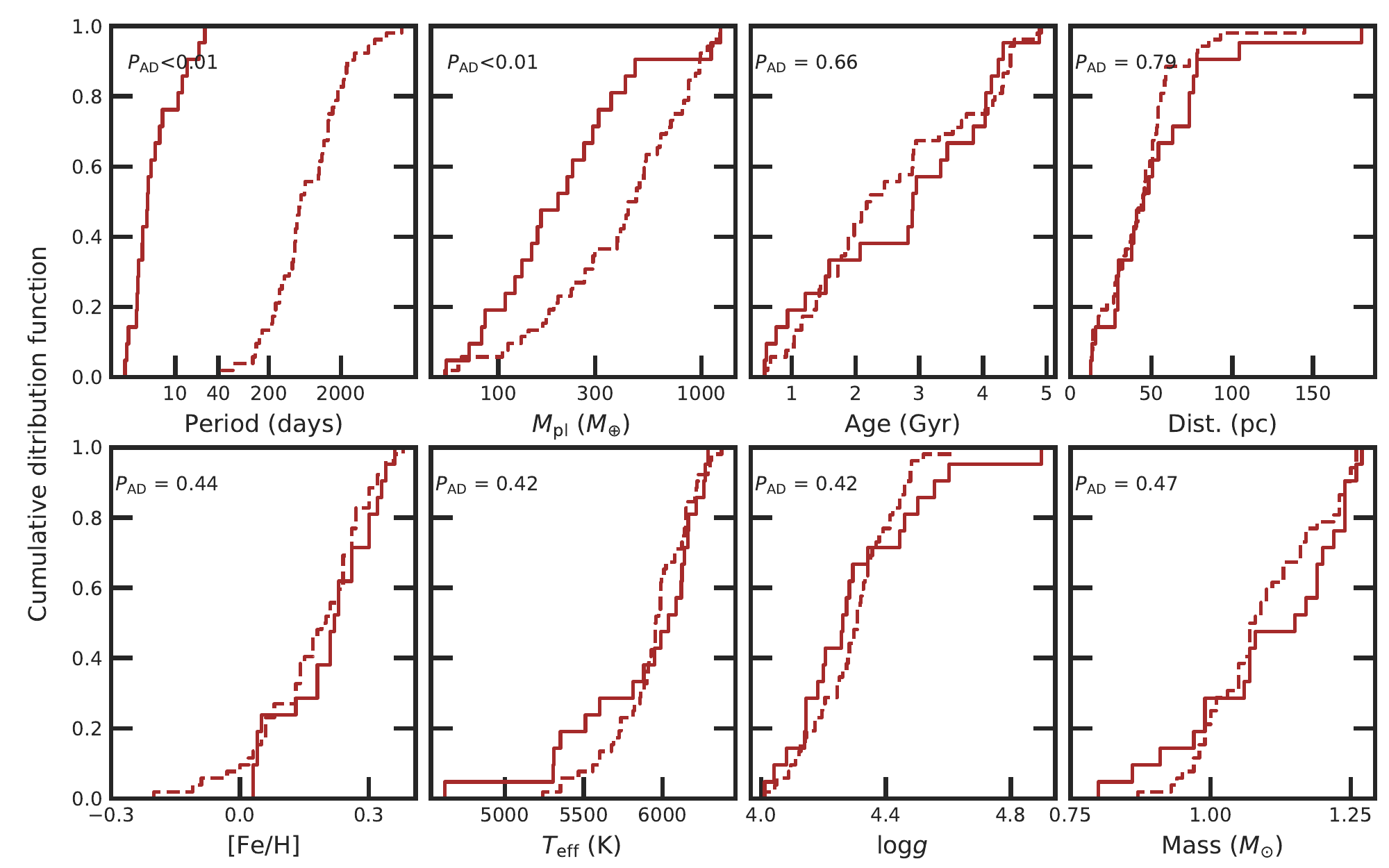}
\end{center}
\vspace{-0.5cm}
\caption{CDFs of different properties of short- (Period $<$ 30 days; solid line) and long-period (Period $>$ 30 days; dashed line) planets and their host stars from the extended FGK$P^{\mathrm{low,high}}$ sample associated with over-densities. The $P_{\mathrm{AD}}$ values for each parameter is shown in the respective plot.}
\label{fig:fgk_young_cumulative_plots_Phigh_short_long}
\end{figure*}

\end{appendix}

\end{document}